\documentclass[twocolumn]{raa}
\usepackage{graphicx,times}             %for PS/EPS graphics inclusion, new
\usepackage{subfig}
\usepackage{dcolumn}% Align table columns on decimal point
\usepackage{bm}% bold math
\usepackage{float}
\usepackage{xcolor}
\usepackage{tabularx}
\usepackage{amsmath}
\usepackage{amssymb}
\usepackage{natbib}
\usepackage{hyperref}
\usepackage{footmisc}
\usepackage{multirow}
\usepackage[figuresright]{rotating}
\begin{document}
   \title{Unit panel nodes detection by CNN on FAST reflector}
  \author{Zhi-Song Zhang\inst{1,2}
  \and Li-Chun Zhu
      \inst{1}
  \and Wei Tang
      \inst{1}
  \and Xin-Yi Li
      \inst{1}
  }
  \institute{National Astronomical Observatories,Chinese Academy of Sciences,Beijing 100012, China; {\it lczhu@bao.ac.cn}\\
        \and
             University of Chinese Academy of Sciences,Beijing 100049, China\\}

  \abstract{
  The 500-meter Aperture Spherical Radio Telescope(FAST) has an active reflector. During the observation, the reflector will deformed into a paraboloid of 300-meters. To improve its surface accuracy, we propose a scheme for photogrammetry to measure the positions of 2226 nodes on the reflector. And the way to detect the nodes in the photos is the key problem in photogrammetry. This paper apply Convolutional Neural Network(CNN) with candidate regions to detect the nodes in the photos. The experiment results show a high recognition rate of 91.5$\%$, which is much higher than the recognition rate of traditional edge detection.
  \keywords{FAST --- photogrammetry --- nodes detect --- convolutional neural network}
  }
\authorrunning{Zhi-Song Zhang, Li-Chun Zhu, Wei Tang, Xin-Yi Li}            %author_head in even pages
\titlerunning{Unit panel nodes detection by CNN}  % title_head in odd pages

\maketitle
\flushbottom

  \section{Introduction}
      FAST (500 - meter aperture spherical radio telescope) is relying on the local characteristic of the karst landscape in Guizhou of China, which is the world's largest single caliber radio telescope. It is one of major infrastructure projects of science and technology in China.

    The FAST active reflector include a reflective surface cable net consisting of nearly 10,000 steel cables with a diameter of 500 meters, reflecting surface unit, hydraulic actuators, ground anchor, and ring beam. Reflector cable net is installed on the lattice type circular ring beam, and it has 2225 connect nodes, through which 4450 reflector units are installed on the cable for reflecting radio waves, and each node of which is connected with the down-tied cables. Each node can be adjusted according to the observation request, thus forming a 300 meter paraboloid in the illumination aperture to complete the observation of the telescope. In order to control the shape of the reflector accurately, the position of all nodes of the entire reflector must be accurately measured in 90 minutes, and the calibration precision index should be below 1.5mm.

    The basic task of photogrammetry is to establish the geometrical relationship between the instantaneous image and the object. Once this relationship is properly restored, we can carefully export the information about the target object from the image (\cite{Atkinson:2003:0031-868X:329}). Photogrammetry is one of the most common methods in telescope reflective surface measurement, which has the advantages of high accuracy and efficiency. By sticking the target point (or projection structure light) on the antenna surface to be tested, one or more cameras are used to take multiple photos of the target point to calculate the surface type of the antenna (\cite{xuwenfeng}). The Arecibo 300-meter telescope in the United States has improved its surface accuracy from 15mm to about 5mm by using photogrammetric method (\cite{Edmundson01photogrammetricmeasurement}). The 13.7 m millimeter-wave radio telescope in Delhi city of Qinghai use photogrammetry method to achieve the accuracy level of 0.083 mm (\cite{fanshenghong}), satisfying the surface precision requirements of millimeter wave observation, and also provides a precise measurement method (\cite{zuoyingxi}) for the study of the gravity deformation.

    The surface measurements mentioned above are all disposable, but the FAST surface measurement is the repeatability measurement, and the maintenance of the targets are very difficult. Based on the requirement of FAST reflection surface measurement, this paper proposes a photogrammetric scheme without target to realize the accurate positioning of the nodes and complete the measurement of reflective surface type precision, through extracting the natural characteristics of the nodes in the photos. The pre-developed photogrammetry equipment Digital Positioning Unit(DPU)(\cite{Hu}) can be used to take photos of the surface on the existing stable foundation pier. The main objective of this paper is to improve the recognition rate of nodes and provide the necessary foundation for the implementation of the photogrammetric scheme.
    %============================= section 2 ===================================

 \section{Programme Design}
\begin{figure}
 \centering
  \includegraphics[width=4in,height=3in]{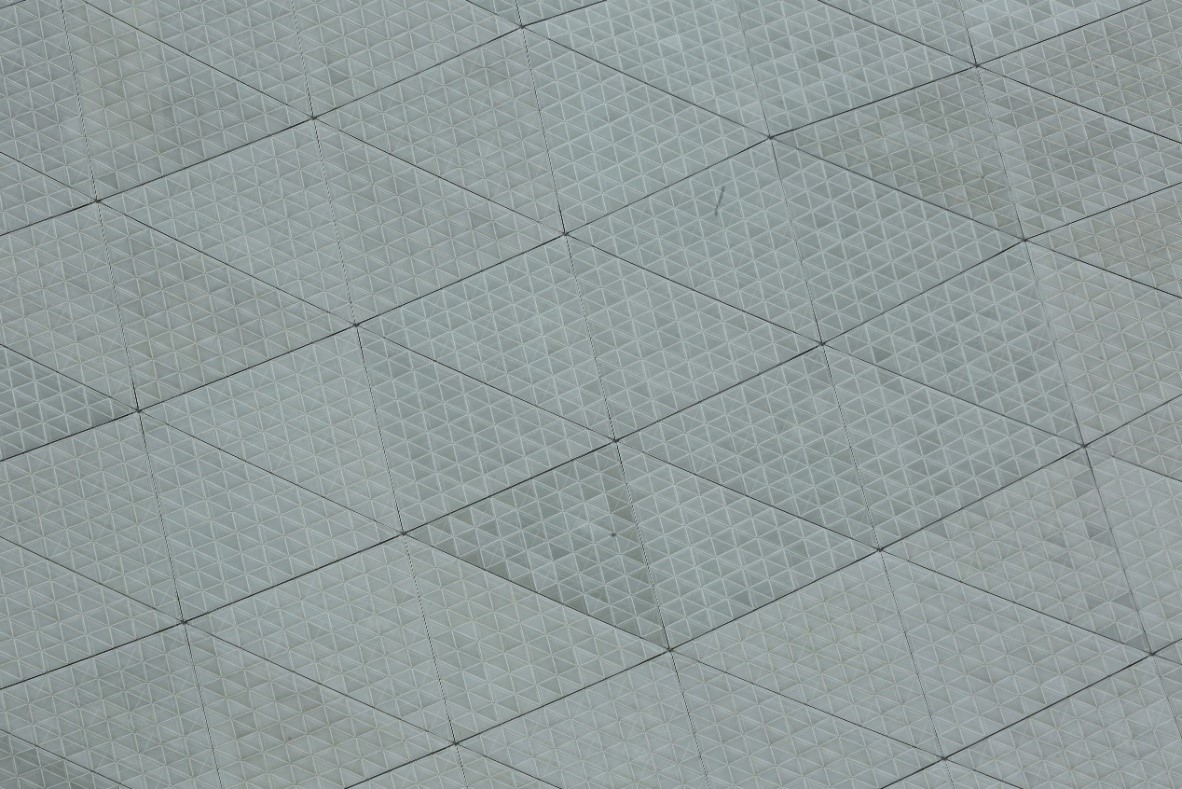}
 \caption{ FAST reflector image  }
 \label{fig:1}
 \end{figure}

\begin{figure}
 \centering
  \includegraphics[width=2in,height=1.5in]{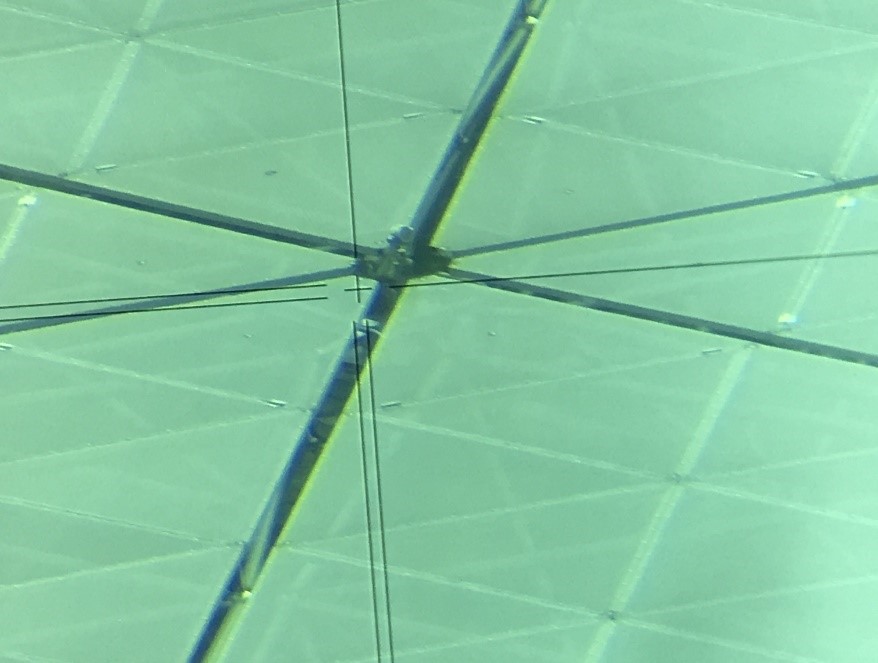}
 \caption{Six gaps around each node, on which are the targets for the total station measurements.}
 \label{fig:2}
 \end{figure}

The FAST reflection plane has unique geometrical features, as shown in Figures \ref{fig:1} , \ref{fig:2}. The reflector is mounted on top of the cable net, which has a height difference of 40cm. The shape of the unit panel is a 11m triangle, and there is a splicing gap between panels. So, this give rise to obvious grey scale contrast between panel and gap when we look over the reflective surface. Cable network node is located at intersection point of the six gaps, which are formed by six panels nearby. This makes it possible to study the measurement method without target. So, this is the basis and core technology to realize the reflecting surface photography measurement without target, using the gap between panels as a feature recognition foundation to find the nodes in photograph.

\begin{figure}
\centering
  \includegraphics[width=4in,height=2in]{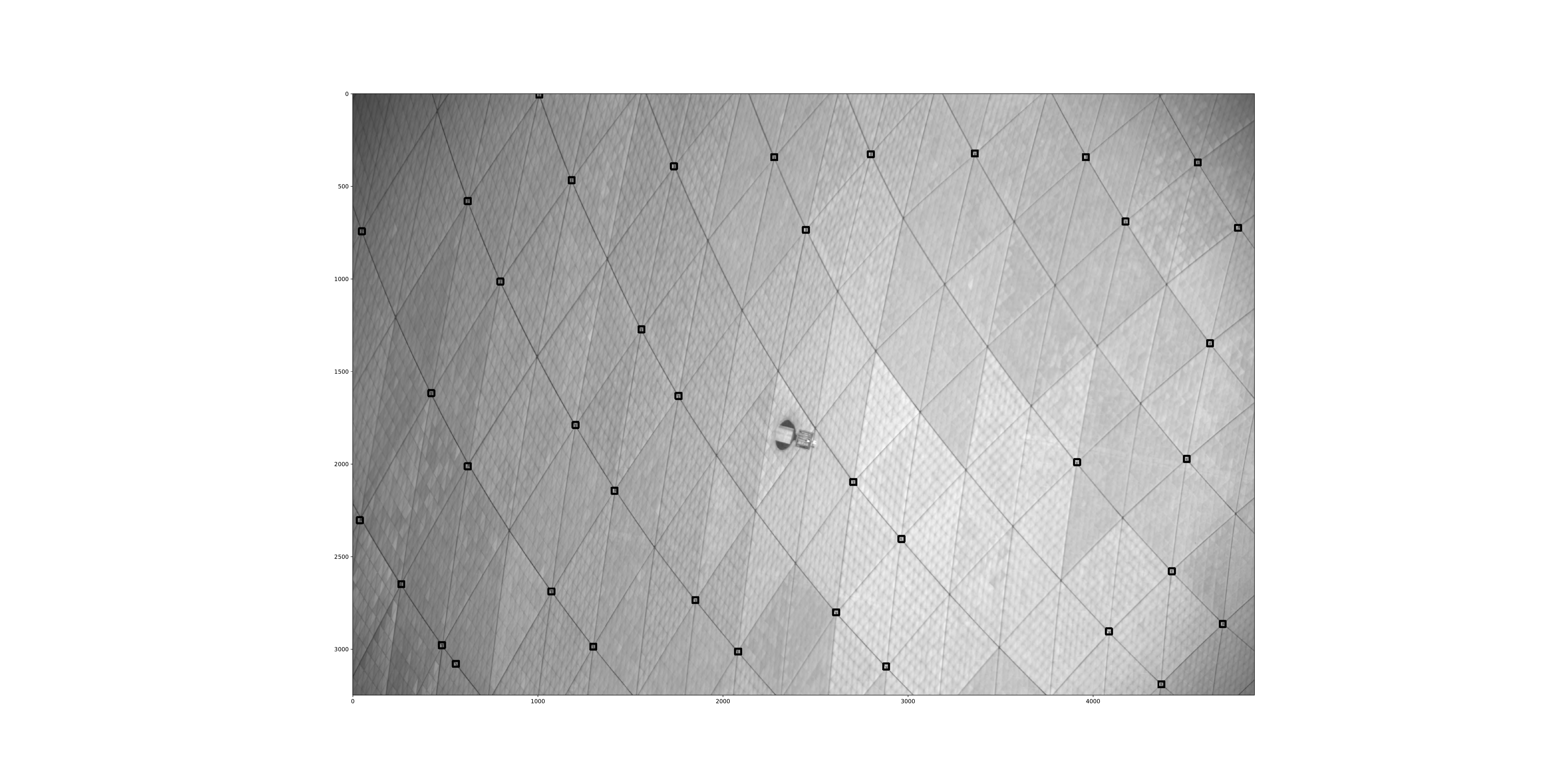}
 \caption{ The recognition of good lighting conditions}
 \label{fig:3}
 \end{figure}

\begin{figure}
 \centering
  \includegraphics[width=4in,height=2in]{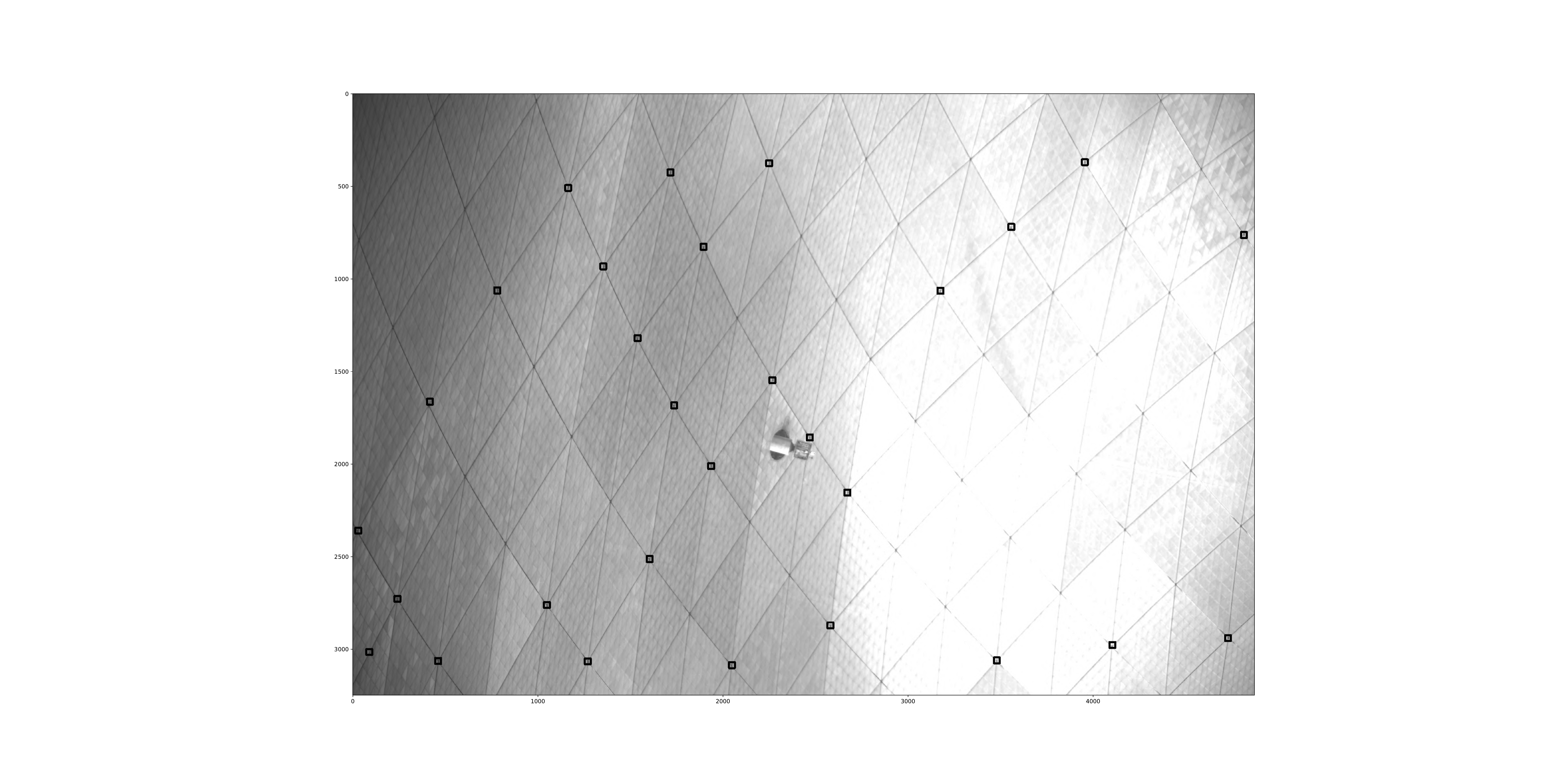}
 \caption{The recognition of uneven or not ideal light conditions}
 \label{fig:4}
 \end{figure}

 At first, we adopt the traditional canny operator edge detection algorithm to identify the nodes (\cite{Canny:1986:CAE:11274.11275}). The experiment result is shown in Figures \ref{fig:3} , \ref{fig:4}. In the case of good lighting conditions, the nodes recognition rate can reach 60$\%$, while the weather changes continually in Guiyang, and often appear cloudy, rainy weather. In the case of uneven or not ideal illumination conditions due to the variable weather, node recognition rate is bad. Considering the changeable weather conditions of FAST, this method cannot achieve a stable high recognition rate.

To solve this problem, we propose a method to identify nodes using Convolution Neural Network(CNN).At present, the nodes recognition of FAST reflector and the characteristics of CNN tally well:

1.	The characteristics of the nodes are highly consistent, which makes the convolutional neural network easier to learn their characteristics.

2.	The recognition target always is the 2225 nodes. This enables us to collect enough sample data to train the network.

Because of the process of modeling, training and testing, CNN is more complex and time-consuming than traditional algorithms. But the measuring equipment and the target image features are fixed, that use the same camera to measure the inherent 2225 nodes.  After the completion of the model training, it is no longer need to change frequently, which greatly improve the practical feasibility of the method.

\begin{figure}
 \centering
 \includegraphics[width=3in,height=3in]{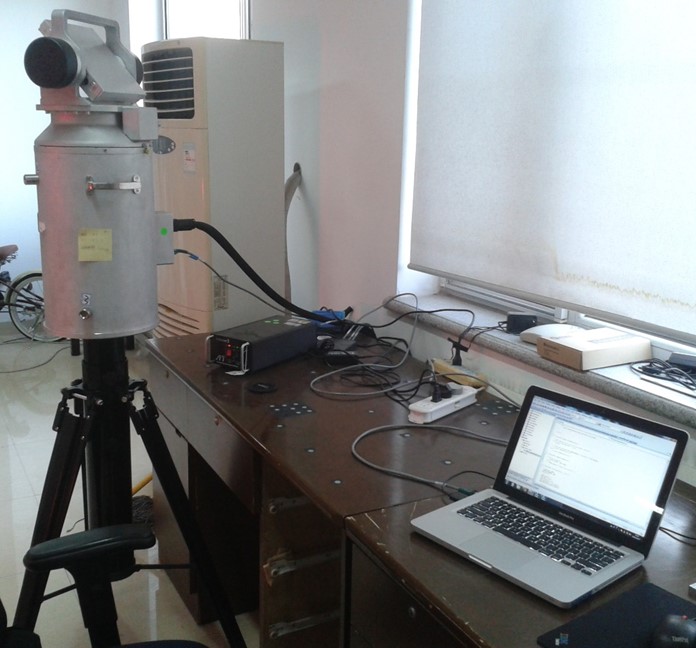}
 \caption{DPU including the body, controller, and control computer}
 \label{fig:5}
 \end{figure}

\begin{figure}
 \centering
 \includegraphics[width=3in]{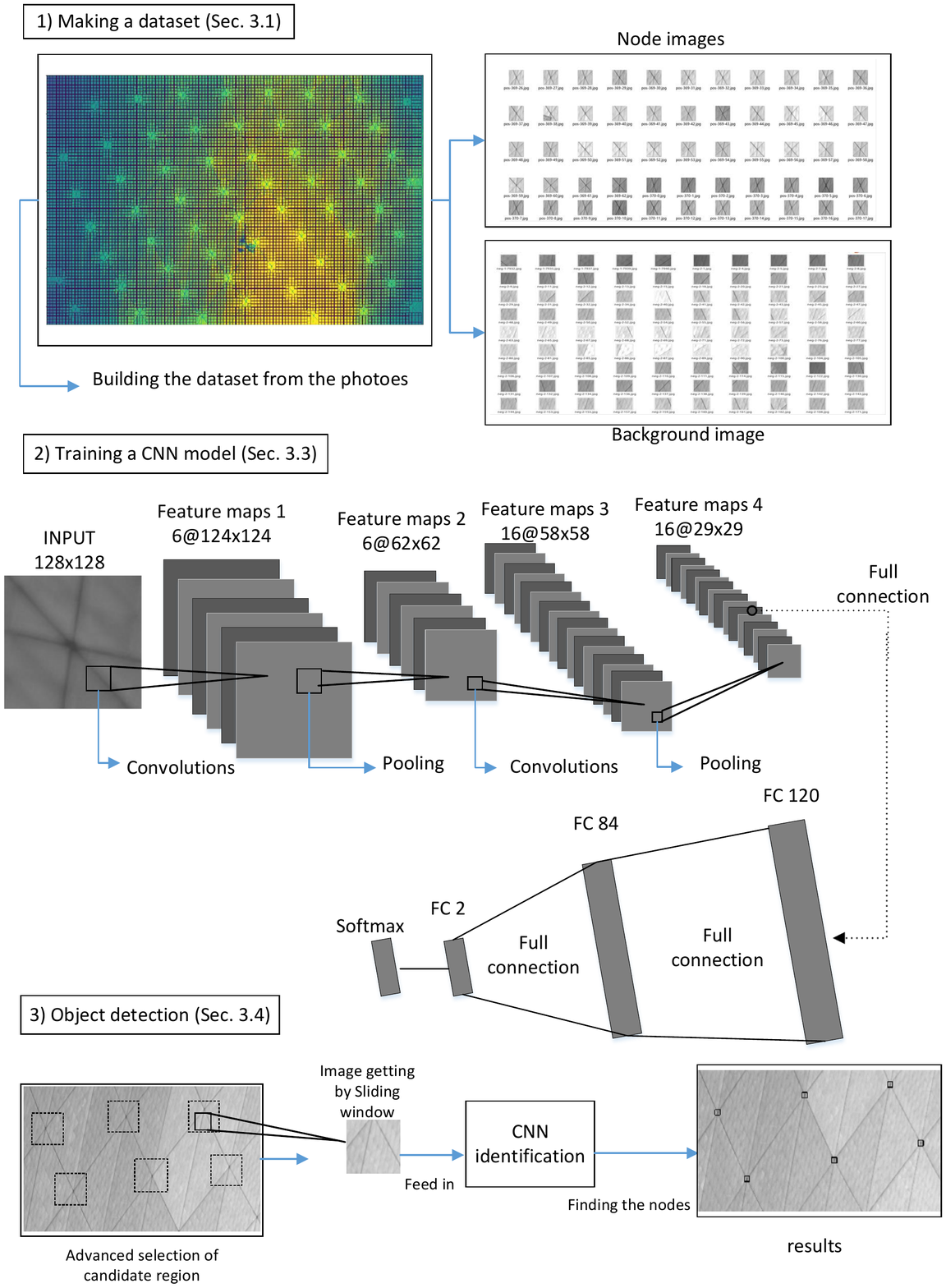}
 \caption{Our proposed approach}
 \label{fig:propose approach}
 \end{figure}

Thanks to the CNN and the region proposal algorithm (\cite{10030193180}), the object detection has made a great breakthrough since 2014. R-CNN (\cite{Girshick:2014:RFH:2679600.2679851}), SPP-NET (\cite{10.1007/978-3-319-10578-9_23}), Fast R-CNN (\cite{DBLP:journals/corr/Girshick15}), Faster R-CNN (\cite{Ren:2015:FRT:2969239.2969250}), SSD (\cite{10.1007/978-3-319-46448-0_2}) and other methods have appeared successively, and these methods have achieved very good results on the online training set. However, due to different tasks and goals, these methods cannot fully meet the specific requirements of FAST node recognition. According to the characteristics of FAST node recognition and on the premise of ensuring the recognition rate and accuracy, a method for object detection with candidate area is proposed in this paper. In the early stage, we have developed a set of special photogrammetric equipment DPU for FAST, as shown in Figure \ref{fig:5}. Combined with the high precision rotary platform in photogrammetry and total station instrument, this equipment can not only achieve high precision and fast measurement, but also meet the requirements of large number of points to be measured and large distribution range. When measuring, the servo motor drives two axes to rotate accurately, so that two or more stereoscopic digital measuring devices are aligned to the measured area at the same time. The focal length of the lens is changed according to the distance between the measured target and the measuring equipment, so as to achieve the target at different distances, and the object surface of the shooting area is the same size. Because there is a big difference in the image of nodes taken in different regions, the model training is carried out in fixed regions in order to eliminate the interference of image differences on the recognition results. In this paper, we conducted experiments on the fixed area near the foundation pier 9 of the FAST, and the selection of the region was controlled by the pitch Angle and azimuth Angle of the DPU. Our proposed approach is shown in Figure \ref{fig:propose approach}.

%============================= section 3 ===================================

\section{Experimental Validation and Date Analysis}

        \subsection{Training Sets Making}
\begin{figure*}
        \centering
        \includegraphics[width=4in,height=2in]{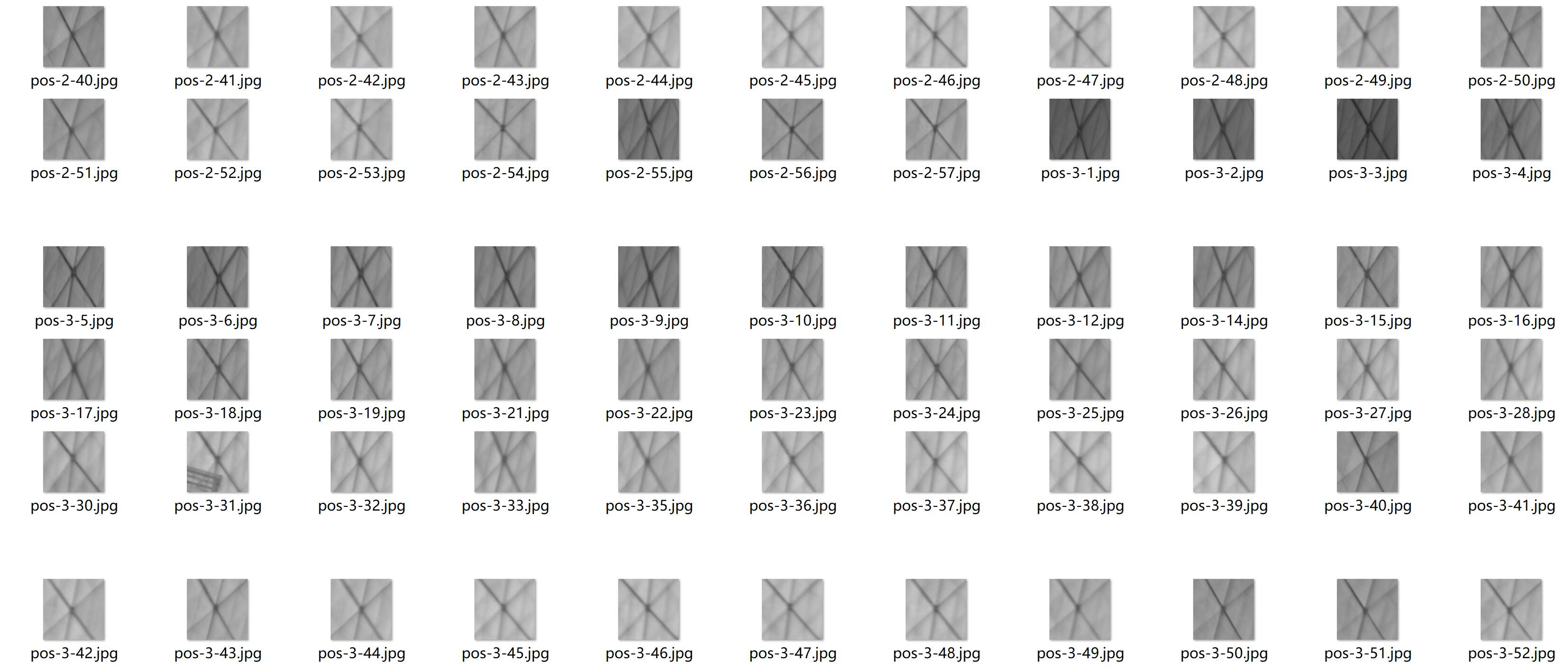}
        \caption{The node training sets}
        \label{fig:6}
        \end{figure*}

Within about seven days in the field of FAST telescope, 460 photoes were taken about the designated area around foundation pier 9 by DPU on foundation pier 6. Each photo has about 60 nodes. The photos includes images in cloudy, sunny, rainy and other light conditions. During the whole shooting process, FAST reflector is in its initial shape. So we calculated the coordinates of the nodes in the photo, which will be introduced in Sec 3.2. We use the node coordinates extracted about 22200 node images from the photos. As shown in Figure \ref{fig:6}.

        Six representative photos were selected, from which about 48,000 background images were extracted by sliding window method.
        The final training set consists of about 22, 000 node images and 48,000 background images.

        \subsection{Nodes Calibration Method}
\begin{figure}
       \centering
       \includegraphics[width=3.5in,height=3in]{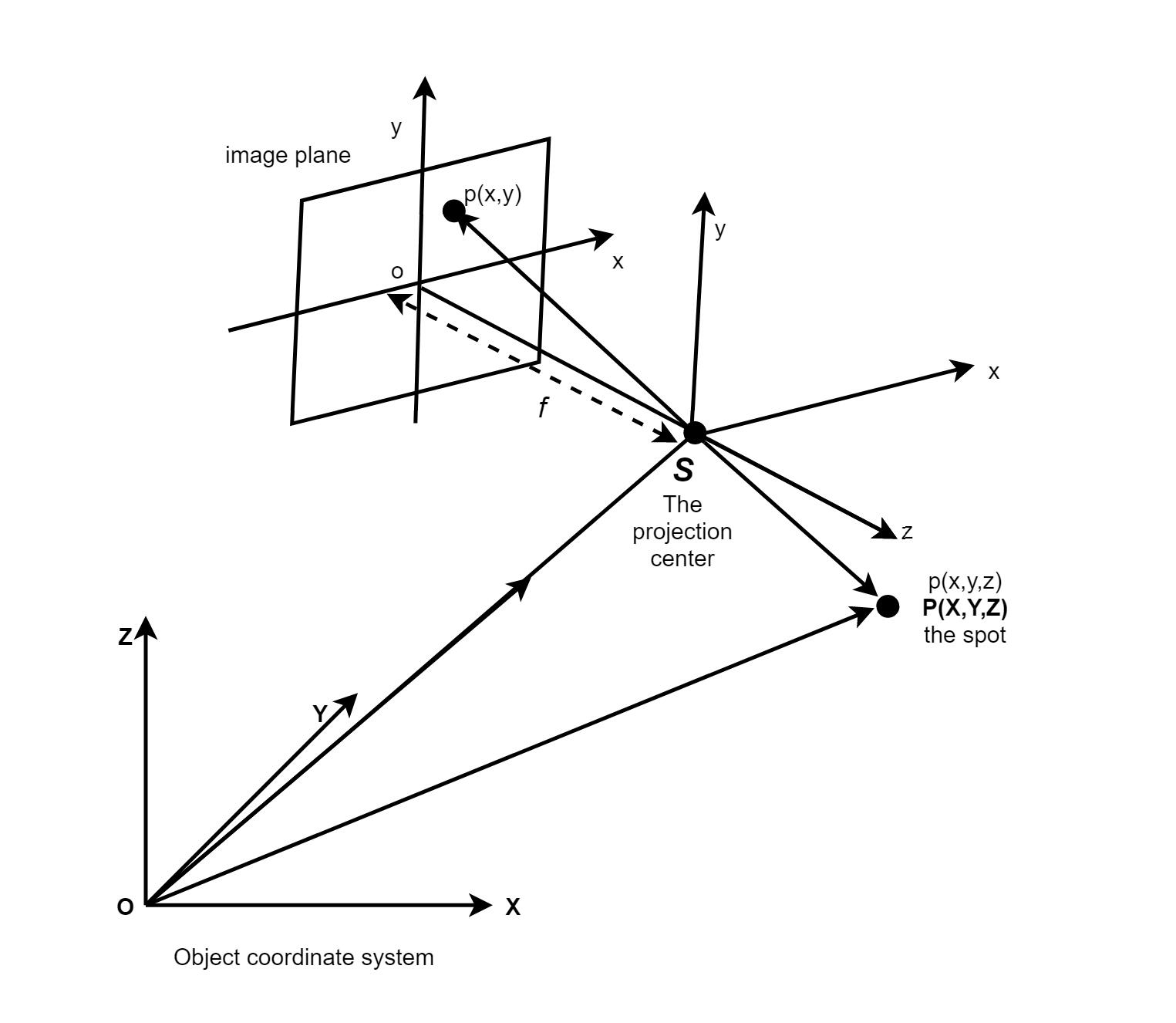}
       \caption{Central perspective projection}
       \label{fig:7}
       \end{figure}

The relationship between nodes on the reflection surface and in the photo are shown in Figures \ref{fig:7}. There are three different coordinates. The object coordinate system is based on the center of FAST reflector, in which the initial coordinates of the nodes are known. Image plane coordinate system is a two-dimensional coordinate system with the photo center as the origin. The projection center coordinate system is obtained by shifting the image coordinate system by a focal distance. The following relationship can be obtained as follows
        \begin{equation}
        \vec{OP}=\vec{OS}+\vec{SP}.
        \label{eq:1}
        \end{equation}

        Let the point P be $\bar{p}$ in the space coordinate system S-xyz, so $\vec{SP}$ can be written as
        \begin{equation}
        \vec{SP}=M\bullet\vec{S\bar{p}},
        \label{eq:2}
        \end{equation}
        where M is the rotation matrix of the projection center coordinate system to the object coordinate system.

        $\vec{Sp}$ is in line with $\vec{S\bar{p}}$ and in the opposite direction, so $\vec{Sp}$ can be written as
        \begin{equation}
        \vec{Sp}=-k\vec{S\bar{p}}(k \rm{~is ~proportional ~coefficient}),
        \label{eq:3}
        \end{equation}
        and put equations(\ref{eq:2}) and (\ref{eq:3}) in formula(\ref{eq:1}), we can find the relationship as follows
        \begin{equation}
        \vec{OP}=\vec{OS}-kM\bullet\vec{Sp},
        \label{eq:5}
        \end{equation}
        and bring in coordinates, formula(\ref{eq:5}) can be written as
        \begin{equation}
        \left[ \begin{array}{lcr}X \\Y \\ Z \end{array}\right]=\left[ \begin{array}{lcr}X_{s} \\Y_{s} \\ Z_{s} \end{array}\right]-k\left[ \begin{array}{lcr}a_{1} & a_{2} & a_{3}  \\b_{1} & b_{2} & b_{3} \\ c_{1} & c_{2} & c_{3}\end{array}\right]\left[ \begin{array}{lcr}x \\y \\ -f \end{array}\right],
        \end{equation}
        where f is the distance from the image plane to the center of photography and \[\left[ \begin{array}{lcr}a_{1} & a_{2} & a_{3}  \\b_{1} & b_{2} & b_{3} \\ c_{1} & c_{2} & c_{3}\end{array}\right]\] is the rotation matrix.

        The inverse transformation is as follows
        \begin{equation}
        \left[ \begin{array}{lcr}x \\y \\ -f \end{array}\right]=-1/k\left[ \begin{array}{lcr}a_{1} & b_{1} & c_{1}  \\a_{2} & b_{2} & c_{2} \\ a_{3} & b_{3} & c_{3}\end{array}\right]\left[ \begin{array}{lcr}X-X_{s} \\Y-Y_{s} \\ Z-Z_{s} \end{array}\right],
        \label{eq:4}
        \end{equation}
       and bring the third subtype of formula (\ref{eq:4}) into the first two subtypes, we can find the relationship as follows
       \begin{equation}
       \left\{\begin{aligned}&x=-f\frac{a_{1}(X-X_{S})+b_{1}(Y-Y_{S})+c_{1}(Z-Z_{s})}{a_{3}(X-X_{S})+b_{3}(Y-Y_{S})+c_{3}(Z-Z_{s})}\\&y=-f\frac{a_{2}(X-X_{S})+b_{2}(Y-Y_{S})+c_{2}(Z-Z_{s})}{a_{3}(X-X_{S})+b_{3}(Y-Y_{S})+c_{3}(Z-Z_{s})}\end{aligned}\right.,
        \label{eq:6}
        \end{equation}

\begin{figure}
        \centering
        \includegraphics[width=3in,height=3in]{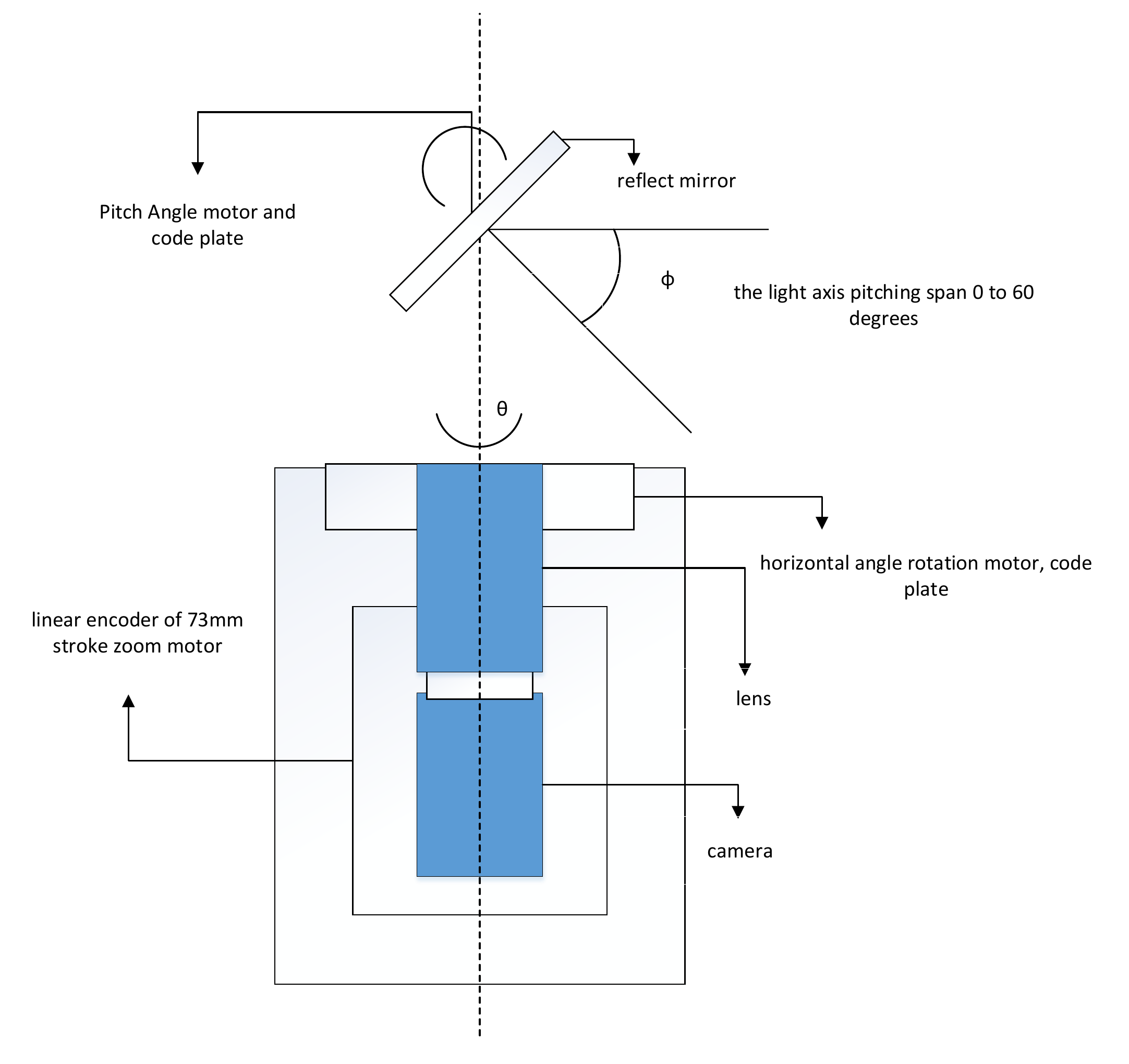}
        \caption{DPU structure}
        \label{fig:8}
        \end{figure}

        The rotation matrix can be written as
        \begin{equation}
        M=\left[ \begin{array}{lcr}a_{1} & a_{2} & a_{3}  \\b_{1} & b_{2} & b_{3} \\ c_{1} & c_{2} & c_{3}\end{array}\right]
        =\left[ \begin{array}{lcr}1 & 0 & 0  \\0 & \cos2\varphi & \sin2\varphi \\ 0 & -\sin2\varphi & \cos2\varphi\end{array}\right]
        \left[ \begin{array}{lcr}\cos\theta & \sin\theta & 0  \\-\sin\theta & \cos\theta & 0 \\ 0 & 0 & 1\end{array}\right]
        =\left[ \begin{array}{lcr}\cos\theta & \sin\theta & 0  \\-\sin\theta\cos2\varphi & \cos\theta\cos2\varphi & \sin2\varphi \\ \sin\theta\sin2\varphi & -\sin2\varphi\cos\theta & \cos2\varphi\end{array}\right],
        \label{eq:7}
        \end{equation}
where \[\left[\begin{array}{lcr}1 & 0 & 0  \\0 & \cos2\varphi & \sin2\varphi \\ 0 & -\sin2\varphi & \cos2\varphi\end{array}\right]\]is the image plane that rotates $2\varphi$  around the x axis, and \[\left[ \begin{array}{lcr}\cos\theta & \sin\theta & 0  \\-\sin\theta & \cos\theta & 0 \\ 0 & 0 & 1\end{array}\right] \] is the image plane that rotates $\theta$  around the z axis, The angle of rotation is determined by the pitch angle and azimuth angle of the DPU in Figure \ref{fig:8}. The pitch angle is doubled because the light is reflected from the reflector to enter the lens.

The position coordinates of the nodes on the image plane are calculated by the coordinates of the projection center and the coordinates of nodes in the object coordinate system through the formula (\ref{eq:6}) and (\ref{eq:7}).

        \subsection{Classifier Training}
\begin{figure*}
        \centering
        \includegraphics[width=4in,height=3in]{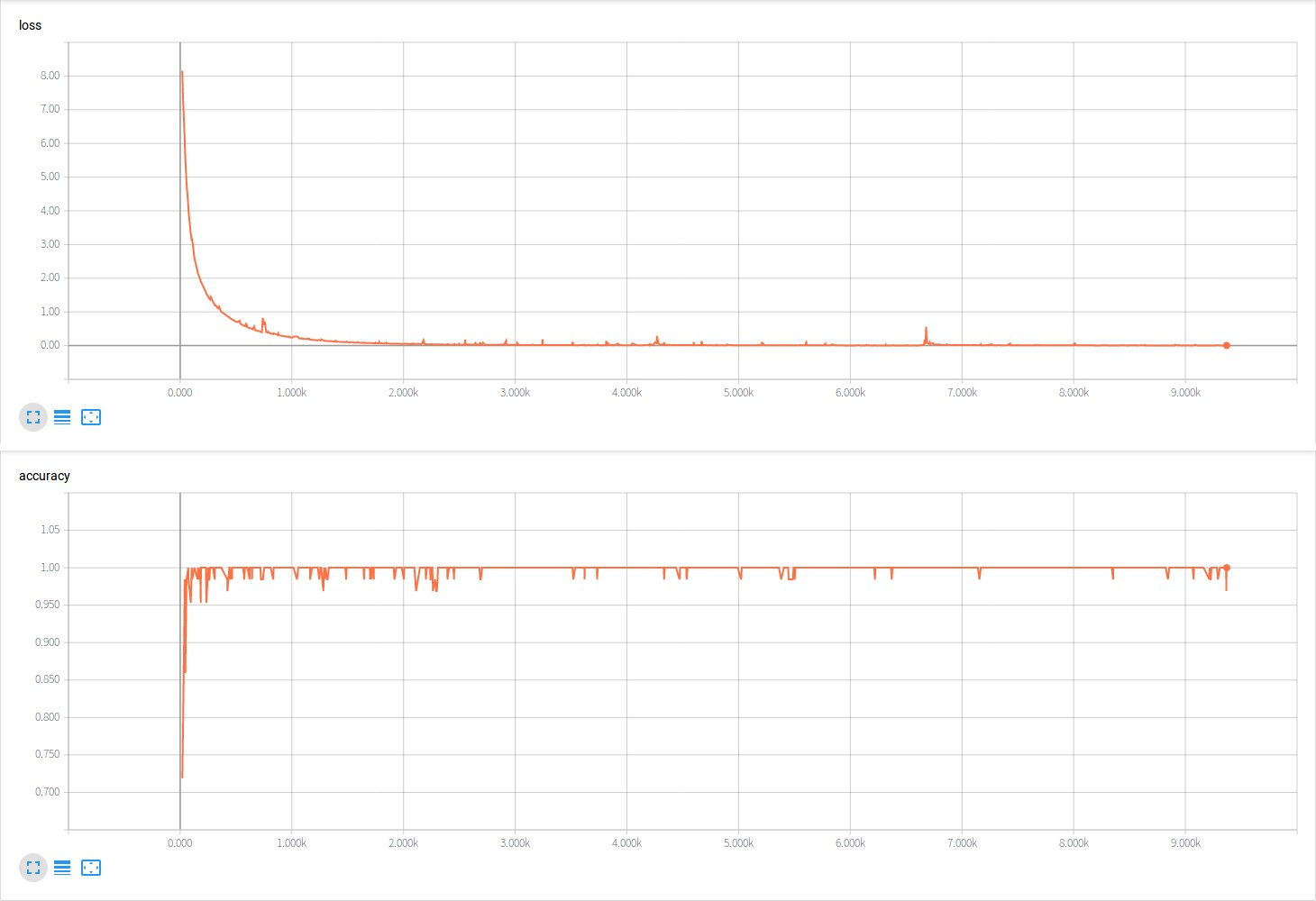}
        \caption{Training results of configuration 1 by Tensorboard}
        \label{fig:9}
        \end{figure*}

\begin{figure*}
        \centering
        \includegraphics[width=4in,height=3in]{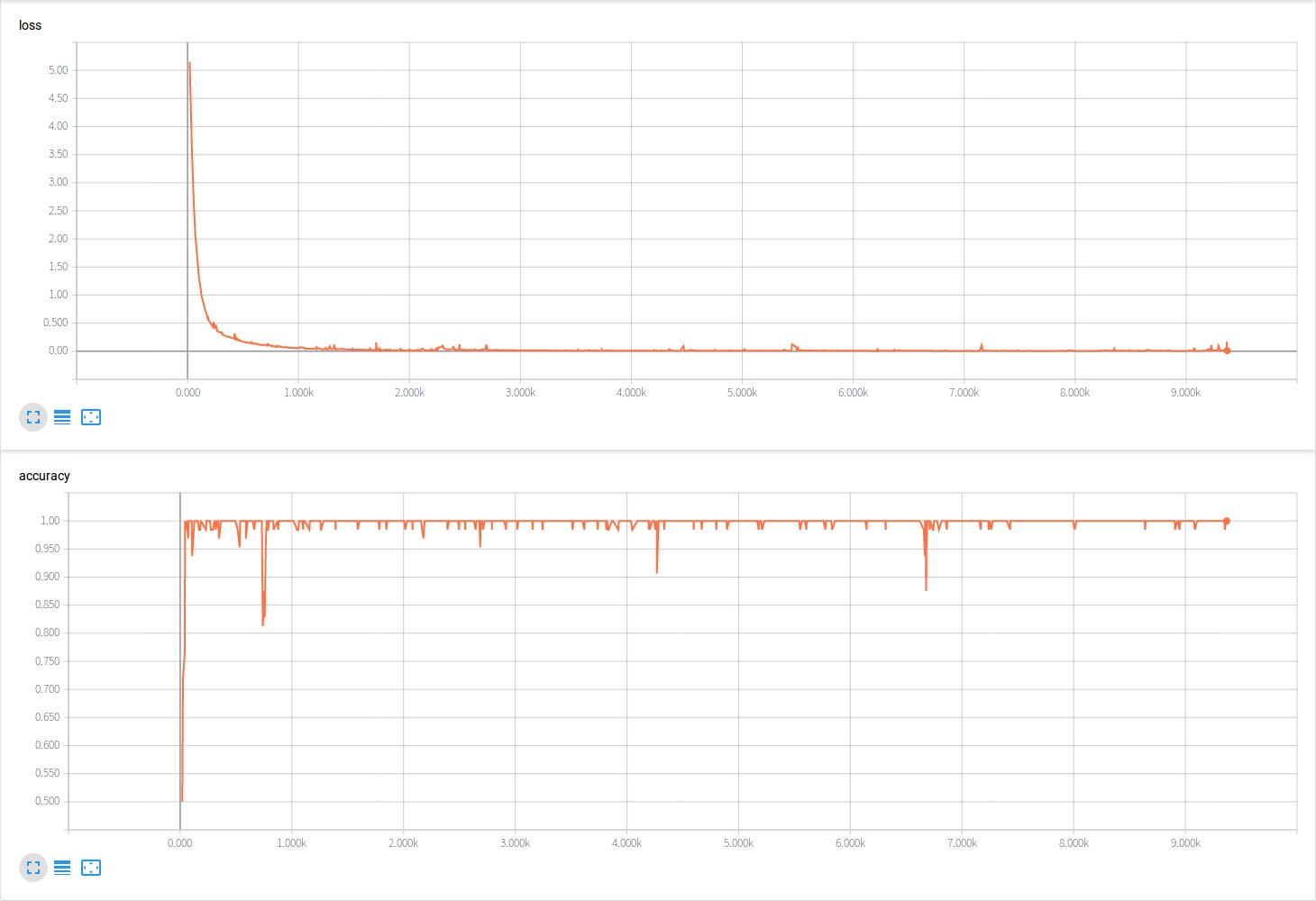}
        \caption{Same as Fig.10, but for configuration 2}
        \label{fig:10}
        \end{figure*}

Since the node image and background image structure are relatively simple, and is grayscale image of 128x128, we select the convolutional network structure, LeNet5 (\cite{5265772}), which is relatively mature for the handwritten experiment. By fine tuning its structure, the input becomes 128x128 matrix, and the output becomes 1x2 vector form. Table \ref{tab:1} shows two convolutional neural network structures. Figures \ref{fig:9}, \ref{fig:10} is the accuracy and loss function of two network structure training processes. It can be seen from the figures that the loss value quickly converges to a small value, but comparing the accuracy, the structure 1 converges better, so we  choose structure 1 finally.

        8,000 images are randomly selected from the more than 60,000 training images to verify the training effect of the model, and the other 50,000 images are trained as training sets. The final accuracy of the model in the training set is 99.89 $\%$, and the final accuracy in the verification set is 99.97 $\%$. Analysis results show that the model has been able to accurately classify nodes and backgrounds in the training set and validation set.

\begin{sidewaystable}[h]
		%\begin{center}
		\resizebox{1\textwidth}{!}{%
			%\begin{tabular}{|P{5mm}|P{3mm}|P{3mm}|P{3mm}|P{3mm}|P{3mm}|P{3mm}|P{3mm}|P{3mm}|P{3mm}|P{3mm}|P{3mm}|P{3mm}|P{3mm}|P{3mm}|P{3mm}|P{3mm}|}
			\begin{tabular}{l|c|c|c|c|c|c|c|c|c|c}
				\hline
				Configuration  \\
				\hline
				
				ConvNet Configuration1 & 5 weight layers & input(128x128x1 image) & conv5-6 & maxpool & conv5-16 & maxpool & FC-120 & FC-84 & FC-2 & ~ \\
				
				ConvNet Configuration2 & 6 weight layers & input(128x128x1 image) & conv5-6 & maxpool & conv5-16 & maxpool & conv5-32 & FC-120 & FC-84 & FC-2  \\
				\hline
			\end{tabular}
		}
		%\end{center}
		\\
		\caption{ConvNet Configurations}
		\label{tab:1}
	\end{sidewaystable}

        \subsection{Advanced Selection of Candidate Regional Methods}At first we attempt to scan the whole picture carried out by sliding window method. The result is not ideal, the node recognition rate is not high, and there are a lot of errors. There are two reasons:

        (1)	The background interference encountered in the scanning process is too much so that it makes the error rate increase.

        (2)	The sliding window slides in 20 pixels, leaving many nodes missing.

        Because DPU¡¯s pitch and azimuth is controlled by providing the node initial position, small candidate areas can be decided around the initial position of the nodes. By reference (\cite{liminghu}) we know that the radial displacement of nodes is less than 0.8 meters. The longitude and latitude displacement¡¯s magnitude even small , we don't need to consider. The candidate areas can be determined as a box of 256x256 pixels around the initial location of the nodes. Then we use the sliding window and non-maximum suppression methods to find object as a result. In order to further correct the position accuracy of the node, the result of the position is shifted up and down with a interval of 1 pixel, then we select the optimal location.

        \subsection{Experimental Results and Analysis}In May 2018, we collected data of the FAST with spherical surface and paraboloidal surface which includes a total of 92 photos. Photos are divided into four categories according to the surface type and light condition, and are detected by the traditional edge detection and the method of this paper respectively. The recognition rate is shown in Table 2 below.

\begin{figure*}
        \centering
          \subfloat[]{%
            \includegraphics[width=3in,height=1.6in]{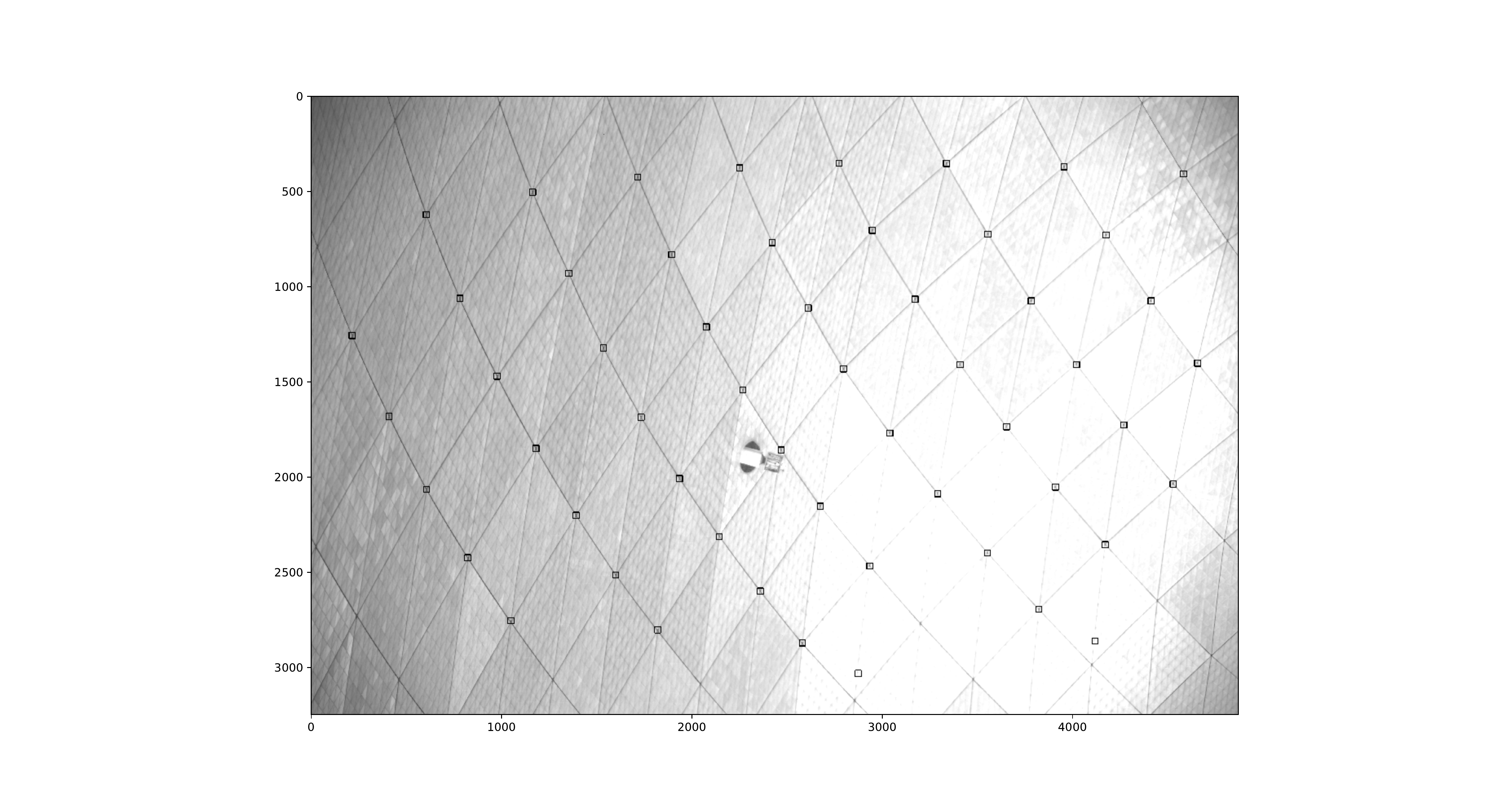}}
          \subfloat[]{%
            \includegraphics[width=3in,height=1.6in]{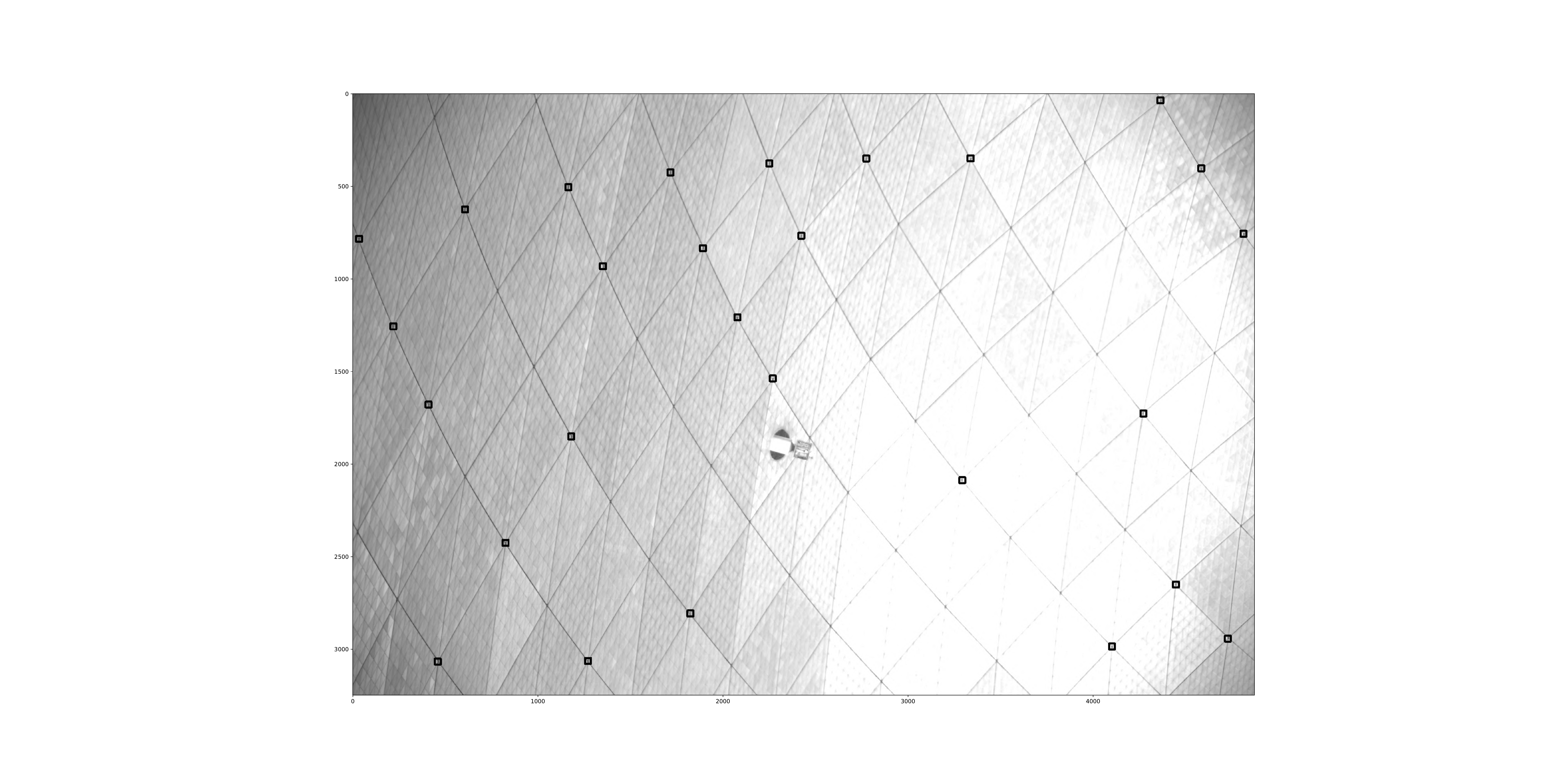}}\\
          \subfloat[]{%
            \includegraphics[width=3in,height=1.6in]{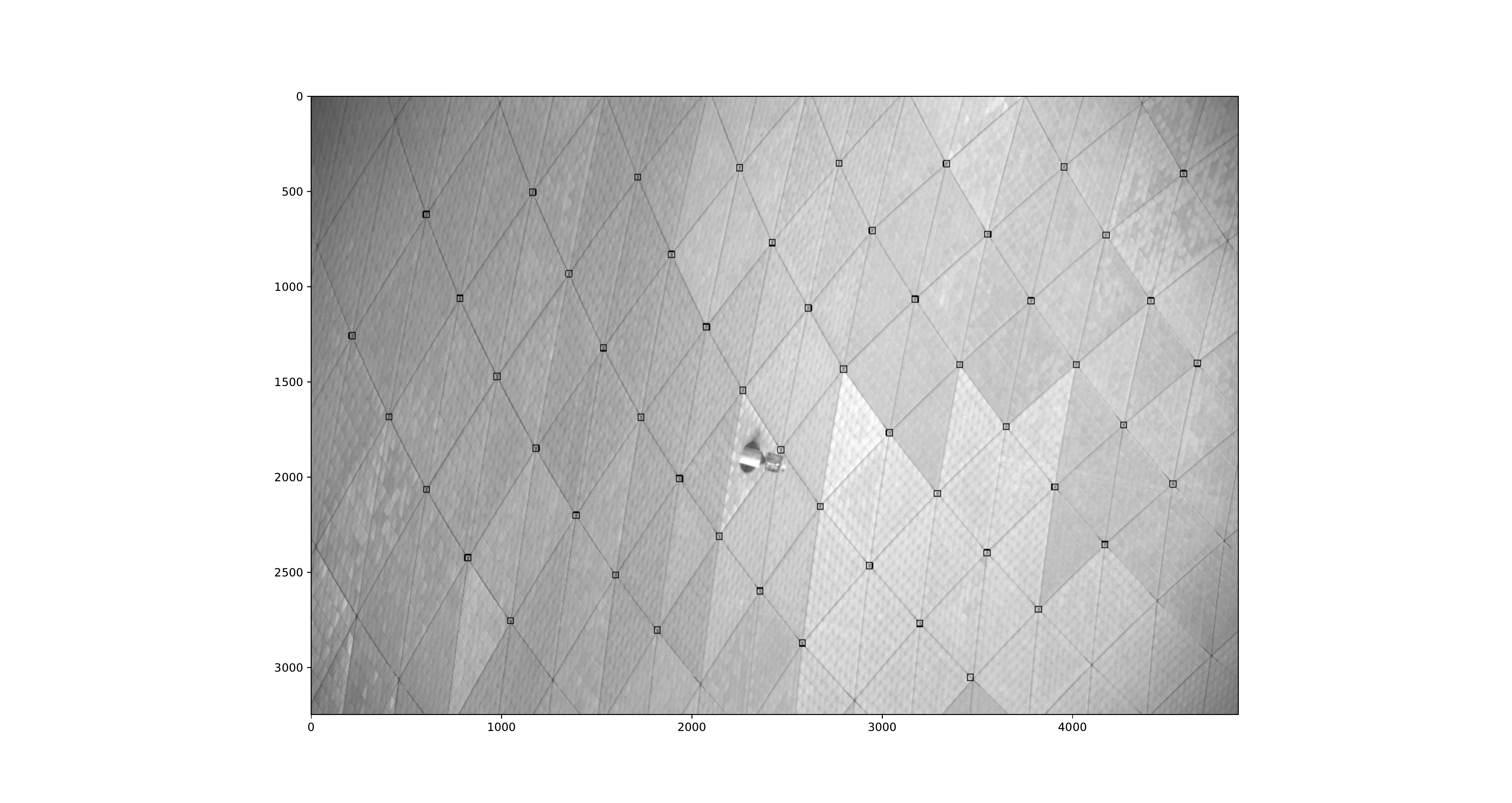}}
          \subfloat[]{%
            \includegraphics[width=3in,height=1.6in]{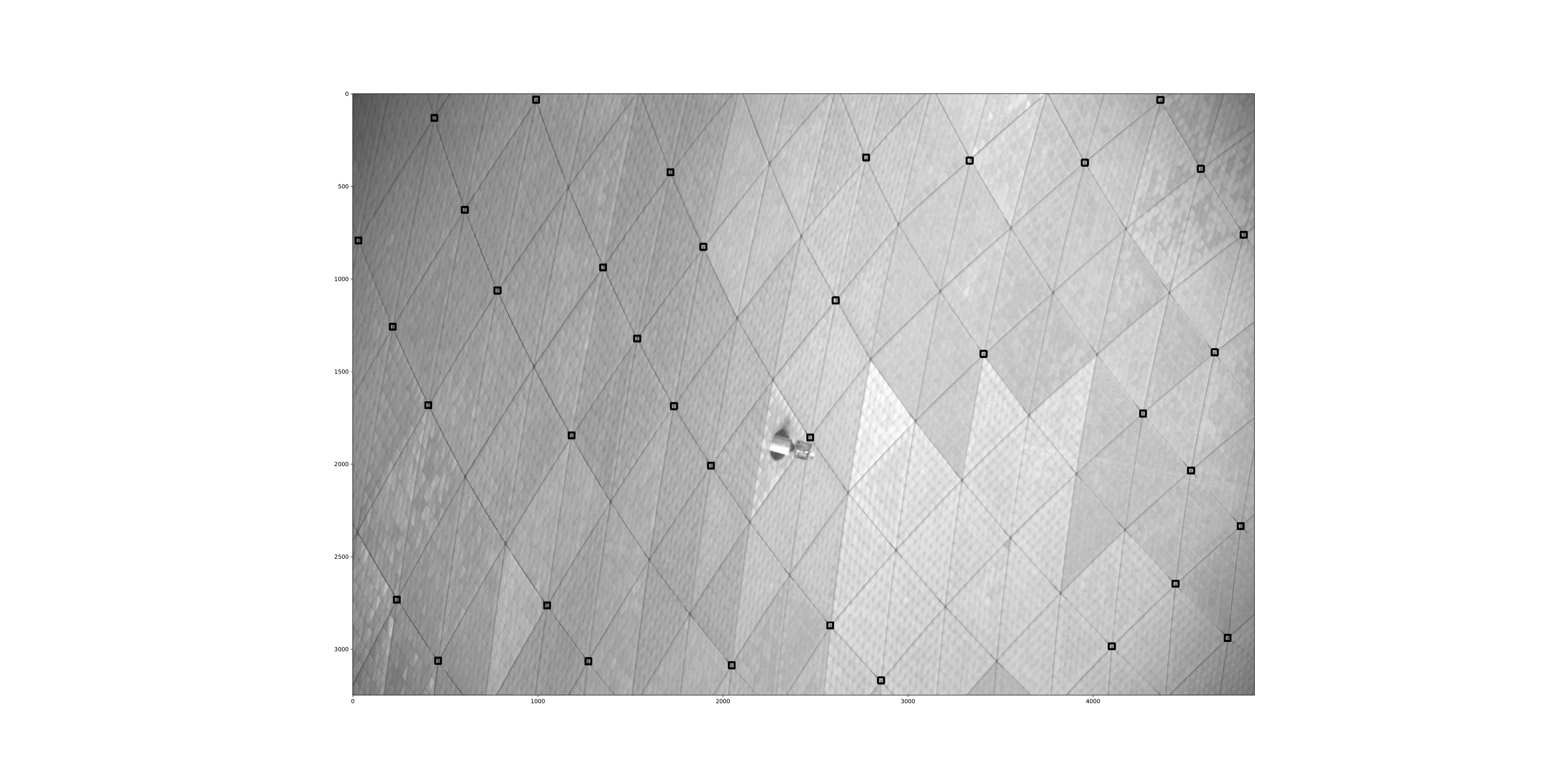}}\\
          \subfloat[]{%
            \includegraphics[width=3in,height=1.6in]{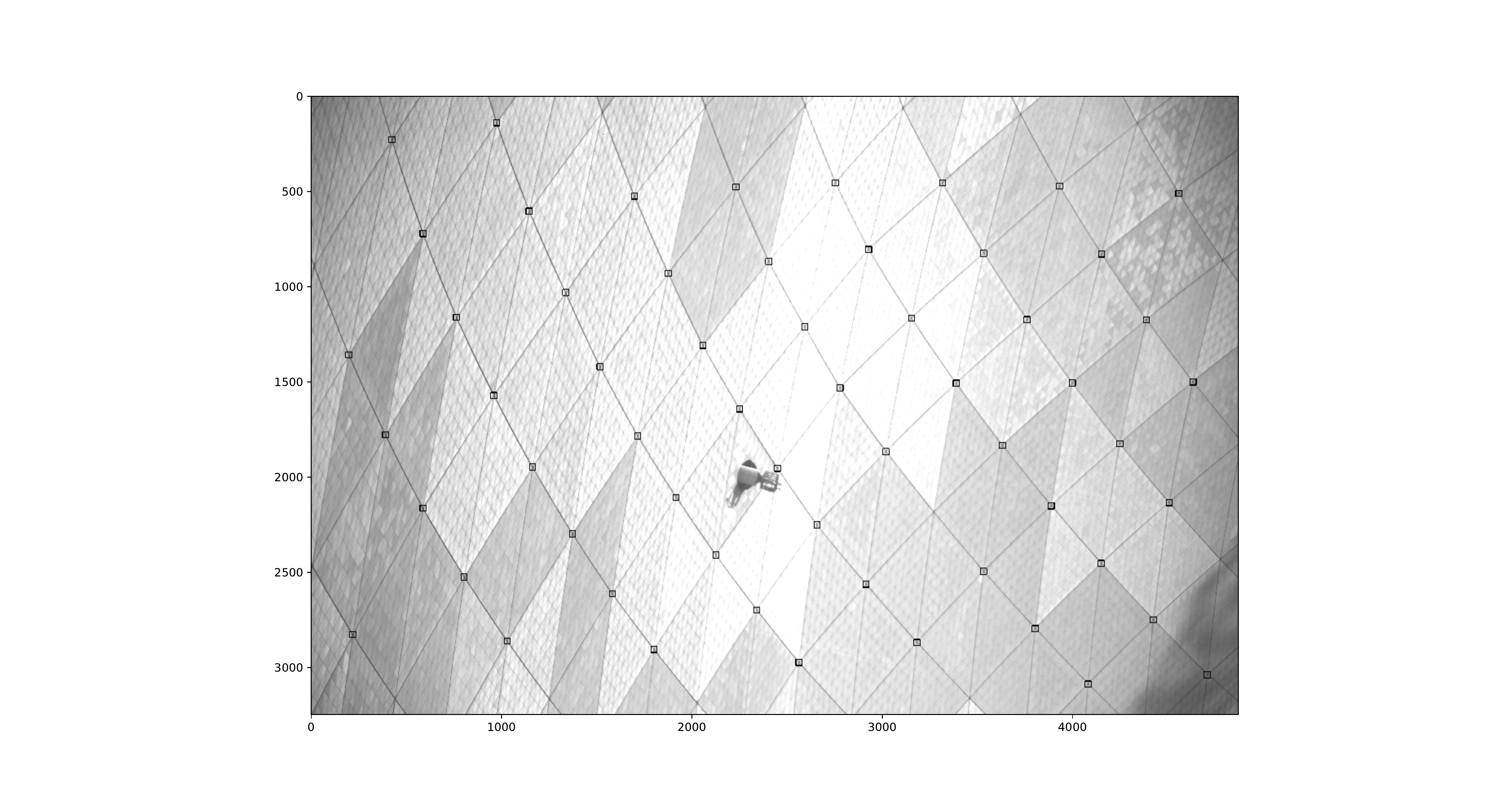}}
          \subfloat[]{%
            \includegraphics[width=3in,height=1.6in]{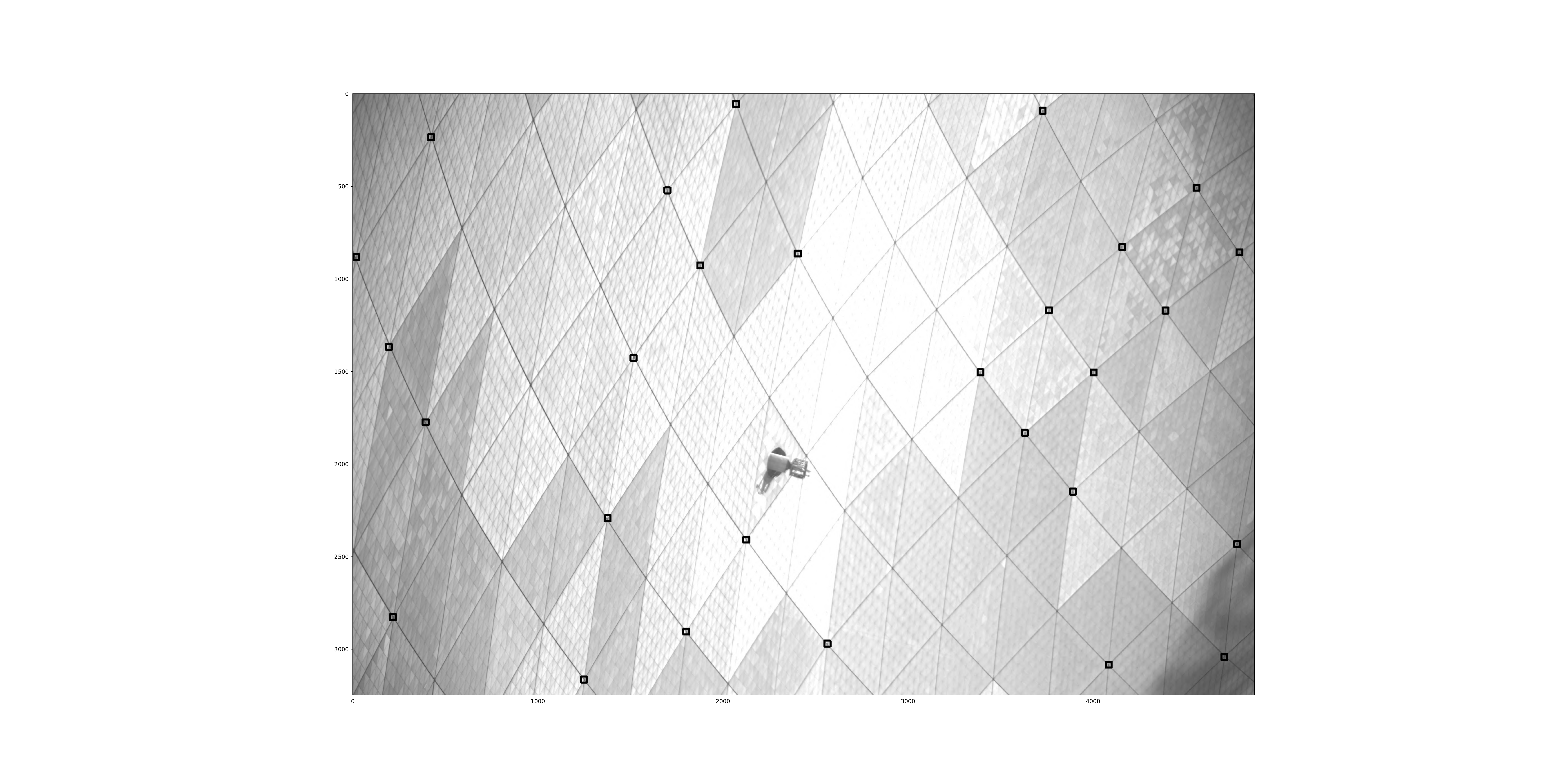}}\\
          \subfloat[]{%
            \includegraphics[width=3in,height=1.6in]{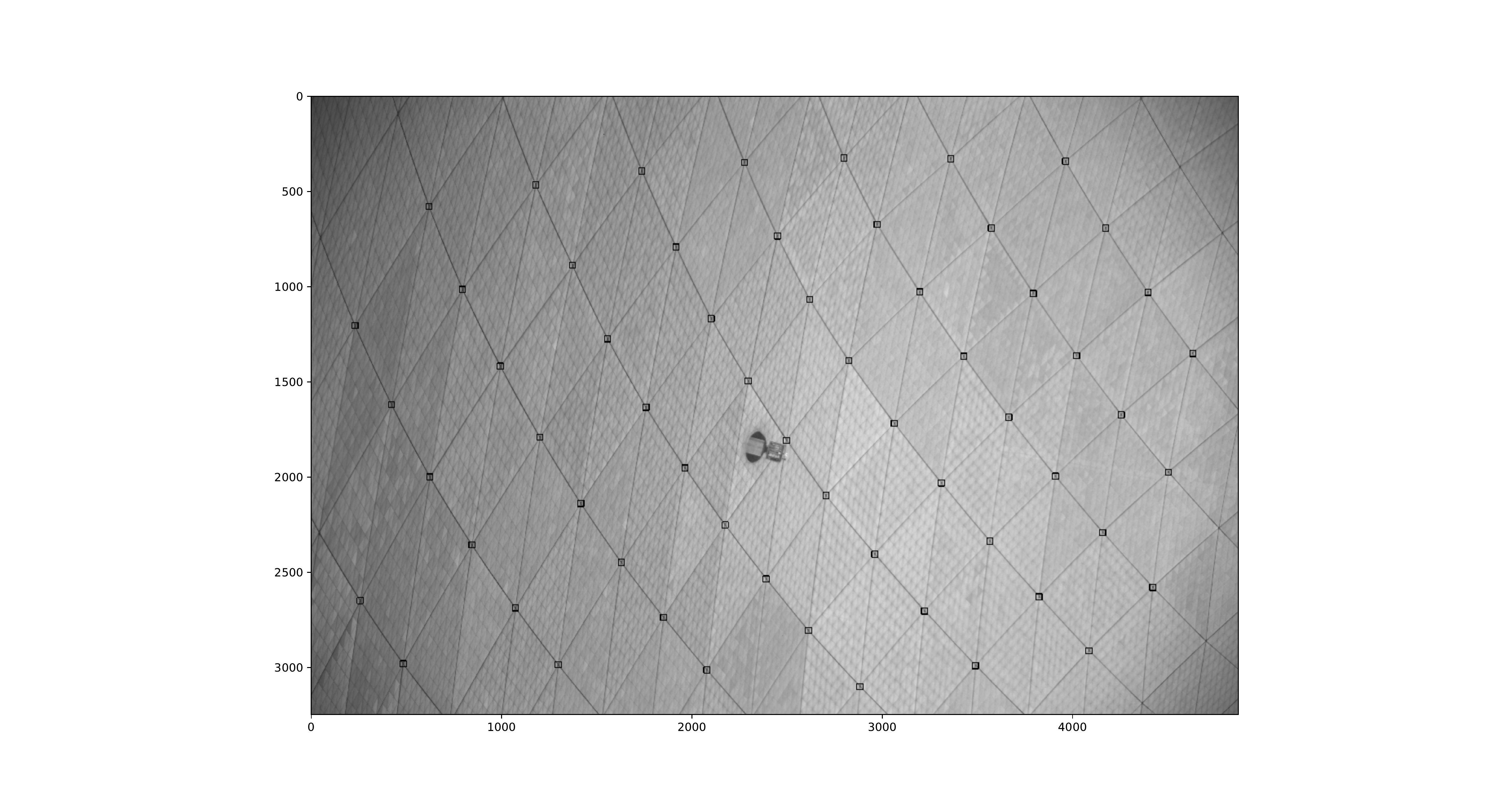}}
          \subfloat[]{%
            \includegraphics[width=3in,height=1.6in]{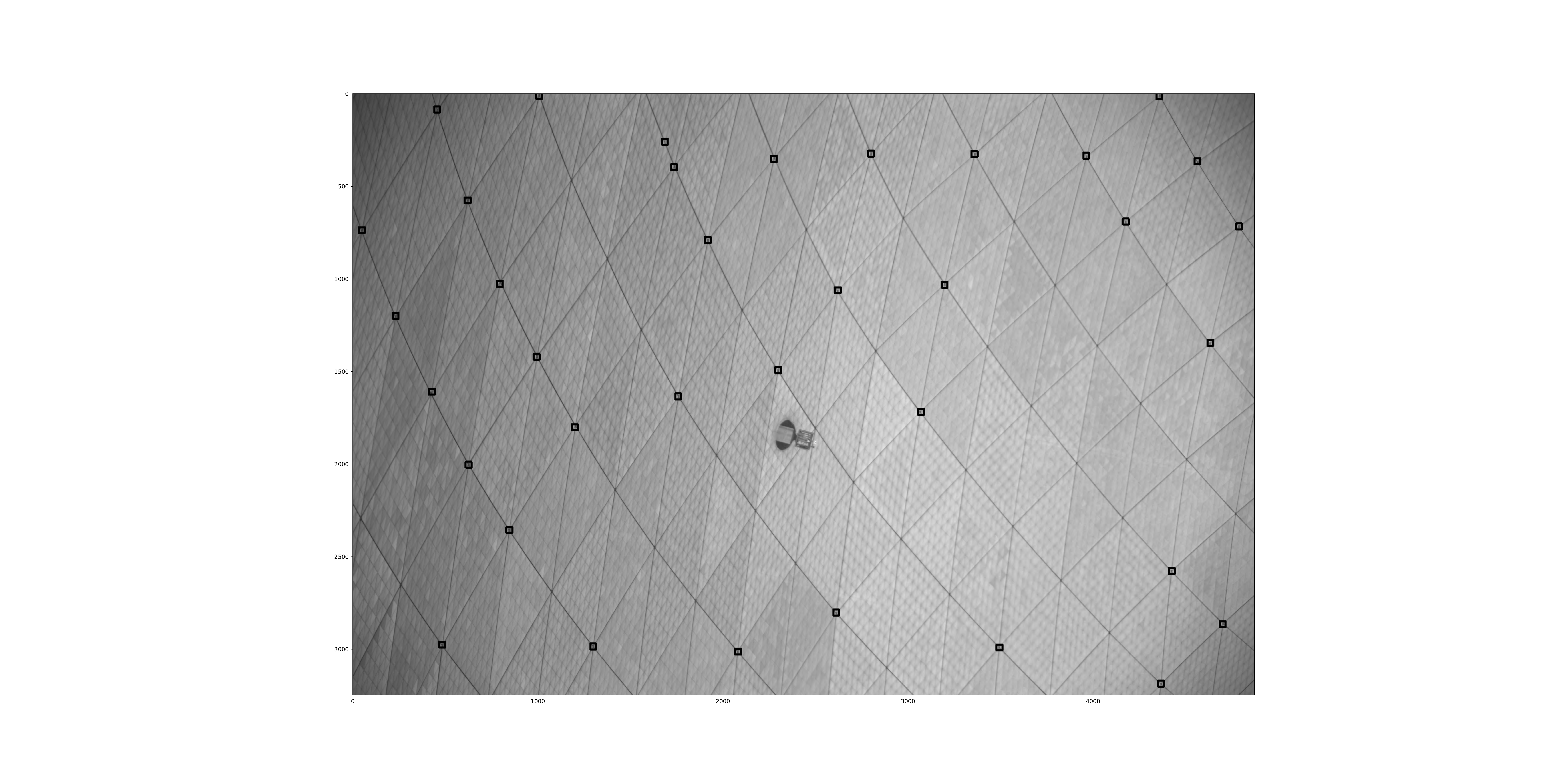}}\\
          \caption{Left column is detected by CNN with Candidate Region, while right column by traditional edge detection. Panels(a),(b)are paraboloid for bad lighting conditions. Panels(c),(d)are paraboloid for good lighting conditions. Panels(e)(f)are sphere for bad lighting conditions. Panels(g)(h)are sphere for good lighting conditions}\label{fig:12}
        \end{figure*}

        Compared with the traditional edge detection, the node recognition rate with this method has a great improvement, and this method has a good robustness to illumination change. The recognition has no significant decline under the light condition of good and bad. The overall recognition rate can reach 91.5$\%$, which meets the requirements of photogrammetry for recognition rate and can be continuously photographed and identified under different lighting conditions. Figure\ref{fig:12} is the contrast between the traditional edge detection method and the effect of this method. We can find that under the condition of various illumination, the recognition rate of this method are better than edge detection method. In addition the effect of our method is very stable.

        \begin{sidewaystable}[h]
		%\begin{center}
		\resizebox{\textwidth}{!}{%
			%\begin{tabular}{|P{5mm}|P{3mm}|P{3mm}|P{3mm}|P{3mm}|P{3mm}|P{3mm}|P{3mm}|P{3mm}|P{3mm}|P{3mm}|P{3mm}|P{3mm}|P{3mm}|P{3mm}|P{3mm}|P{3mm}|}
			\begin{tabular}{l|c|c|c|c|c|c|c}
				\hline
				Methods & sphere for all & sphere for good condition & sphere for bad condition & paraboloid for all & paraboloid for good condition & paraboloid for bad condition & total\\
				\hline
				
				Traditional Edge Detection & 52.1$\%$ & 56.2$\%$ & 47.6$\%$ & 51.3$\%$ & 56.9$\%$ & 44.9$\%$ & 51.5$\%$ \\
				
				CNN with Candidate Region & 95.5$\%$ & 96.8$\%$ & 93.6$\%$ & 89.4$\%$ & 90.7$\%$ & 86.8$\%$ & 91.1$\%$ \\
				\hline
			\end{tabular}
		}
		%\end{center}
		\\
		\caption{Recognition rate for the two methods}
		\label{tab:2}
	    \end{sidewaystable}

%============================= section 3 ===================================

\section{Discussions and Conclusions}

In recent years, object detection has developed rapidly, and many innovative methods have been proposed, but these methods are mostly based on the online training set. According to the actual work requirement, we apply the object detection method to FAST reflecting surface nodes identification. Considering the actual conditions, this paper propose a method of the training sets making and a method of choosing the candidate regions, and successfully improve the classification accuracy of network to 99.9$\%$, and node recognition rate increase to 91.5$\%$. The experiments show that this method is better than the traditional edge detection method, and lays a foundation for the feasibility of the whole non-target photogrammetry scheme. Node identification is a key part of photogrammetry. Next, we will calibrate the DPU and camera, and finally get the position of the node in the object coordinate system through the joint calculation of two DPU.

\section*{Acknowledgments}
This work was supported by study on the fusion of Total station dynamic tracking measuring and IMU inertial measuring for the feed support measurement in FAST(Grant No.11503048) and the Open Project Program of the Key Laboratory of FAST, NAOC, Chinese Academy of Sciences and the Key Laboratory of Radio Astronomy, Chinese Academy of Sciences

\bibliographystyle{raa}
\bibliography{mine}

\end{document}